\documentclass[onecolumn,showpacs]{revtex4}
\usepackage{amsmath,amssymb,graphicx}
\usepackage{color,ulem}

\begin{document}
\title{Weighted Fractal Networks}
\author{Timoteo Carletti$^*$, Simone
  Righi} 
\affiliation{D\'epartement de Math\'ematique, Facult\'es Universitaires
  Notre Dame de la Paix\\ 8 rempart de la vierge B5000 Namur, Belgium\\
corresponding author (*) timoteo.carletti@fundp.ac.be\\
tel. +32(0)81724903\quad fax. +32(0)81724914}
\date{\today}

\begin{abstract}
In this paper we define a new class of weighted complex
    networks sharing several properties with fractal sets, and whose topology
    can be 
  completely analytically characterized in terms of the involved
  parameters and of the fractal dimension. The proposed framework defines an
  unifying general theory of fractal networks able to unravel some hidden
  mechanisms responsible for the emergence
  of fractal structures in Nature.
\end{abstract}

\maketitle

\section{Introduction}

Complex networks have recently attracted a growing interest of 
scientists from different fields of research, mainly because complex networks
define a
powerful framework for describing, analyzing and modeling real systems that
can be found in Nature and/or society. This framework allows
to conjugate the micro to the macro abstraction levels: nodes can be endowed
with local dynamical rules, 
while the whole network can be though to be composed by hierarchies of clusters
of nodes, that thus exhibits aggregated behavior.

The birth of graph theory is usually attributed to L. Euler with his seminal
paper concerning the \lq\lq K\"onigsberg bridge 
problem\rq\rq (1736), but it is only 
in the 50's that network theory started to develop autonomously with the
pioneering works 
of Erd\H os and R\'enyi~\cite{ErdosReny1959}. Nowaday network theory defines a
research 
field in its own~\cite{AB2002,BLMCH2006} and the scientific activity is mainly
devoted to construct and characterize complex networks exhibiting some of the
remarkable properties of real networks, scale--free~\cite{BA1999},
small--world~\cite{WattsStrogatz1998}, communities~\cite{Fortunato2009}, 
just to mention few of them. 

In a series of recent papers~\cite{Zhang2008,Zhang2008EPJB65,Guan2009} authors
proposed a new point of view by constructing networks exhibiting scale-free
structures following ideas taken from fractal construction, e.g. Koch curve or
Sierpinski gasket. The aim of the present paper is to generalize these latter
constructions and to define a unifying theory, hereby named {\it Weighted
  Fractal Networks}, WFN for short, whose networks share with fractal sets
several interesting properties, for instance the self-similarity.

The WFN are constructed via an explicit algorithm and we are
able to completely analytically characterize their topology as a function of
the parameters involved in the construction. We are thus able to prove that
WFN exhibit the \lq\lq small--world\rq\rq property, i.e. slow (logarithmic)
increase of the average shortest path with the network size, and large average
clustering coefficient. Moreover the probability distribution of node strength
follows a power law whose exponent is the Hausdorff (fractal) dimension
of the \lq\lq underlying\rq\rq fractal, hence the WFN are scale--free.

WFN also represent an explicitely computable model for the renormalization
procedure recently applied to complex
networks~\cite{SongHavlinMakse2005,SongHavlinMakse2006,RBFR2008}.  

The paper is organized as follows. In the next section we will introduce the
model and we outline the similarities with fractal
sets. In Section~\ref{sect:result} we present the analytical
characterization of such networks also supported by dedicated numerical
simulations. We then introduce in Section~\ref{sec:nhwfn} a straightforward
generalization of the previous theory, and thus we conclude by showing a
possible application of 
WFN to the study of fractal structures emerging in Nature.


\section{The model}
\label{sect:model}

According to Mandelbrot~\cite{Mandelbrot1982} \lq\lq a fractal is by
definition a set for which the Hausdorff dimension strictly
exceeds the topological dimension\rq\rq. One of the most amazing and
interesting feature of fractals is their {\it self-similarity}, namely looking
at all scales we can find conformal copies of the whole set.
Starting from this property one can provide rules to build up
fractals as fixed point of {\it Iterated
  Function Systems}~\cite{barnsley1988,edgar1990}, IFS for short, whose
Hausdorff dimension is completely characterized by
two main parameters, the number of copies $s>1$ and the scaling factor $0<f<1$
of 
the 
IFS. Let us observe that in this case this dimension coincides with the so  
called similarity dimension~\cite{edgar1990}, $d_{fract}=-\log s/\log f$.

The main goal of this paper is to generalize such ideas to networks, aimed at
constructing  weighted complex
networks~\footnote{We hereby present the construction for undirected networks,
  but it can be straightforwardly generalized to
  directed graphs as well.} with some a priori prescribed topology, that will be described in
terms of node strength distribution, average (weighted) shortest path and
average (weighted) clustering coefficient, depending on the two
main parameters: the number of copies and the scaling factor~\footnote{A
  straightforward generalization will be presented in the next
  Section~\ref{sec:nhwfn}.  See also~\cite{Carletti2009} where the WFN theory
  will be  
  generalized as to include a stochastic iteration process.}. Moreover taking advantage of
the similarity 
with the IFS fractals, some topological properties of the networks will depend
on the fractal dimension of the IFS fractal.

\begin{figure}
\centering
\includegraphics[width=8cm]{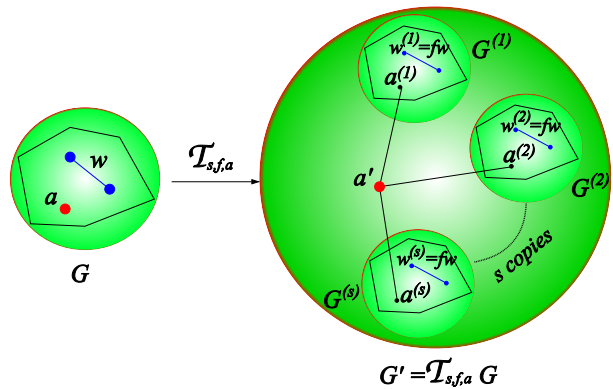}
\caption{The definition of the map $\mathcal{T}_{s,f,a}$. 
On the left a generic initial graph $G$ with its attaching node $a$ (red
on-line) 
and a generic weighted edge $w\in G$ (blue on-line). On the right the new
graph $G^{\prime}$ obtained as follows: Let $G^{(1)},\dots,G^{(s)}$ be $s$ copies of $G$,
whose weighted edges (blue on-line) have
 been scaled by a factor $f$. For $i=1,\dots,s$ let us
 denote by $a^{(i)}$ the node in $G^{(i)}$ image of the labeled
 node $a\in 
 G$, then link all those labeled nodes to a new node $a^{\prime}$ (red
 on-line) through
 edges of unitary weight. The connected
 network obtained linking the $s$ copies $G^{(i)}$ to the node $a^{\prime}$
 will be by definition the image of $G$ through the map: 
 $G^{\prime}=\mathcal{T}_{s,f,a}(G)$.}
\label{fig:construction}
\end{figure}

Let us fix a positive real number $f<1$ and a positive integer $s >1$ and let
us consider a (possibly) weighted network $G$ composed by $N$ 
nodes, one of which has been labeled {\it attaching node} and denoted by
$a$. We then define a map,
$\mathcal{T}_{s,f,a}$, depending on the two parameters $s$, $f$ and on the
labeled node $a$, whose action on networks is described in
Fig.~\ref{fig:construction}. 

So starting with a given initial network $G_0$ we can construct a family of
weighted networks 
$(G_k)_{k\geq 0}$ iteratively applying the previously defined map:
$G_k:=\mathcal{T}_{s,f,a}(G_{k-1})$. 

\begin{figure*}
\centering
\includegraphics[width=18cm]{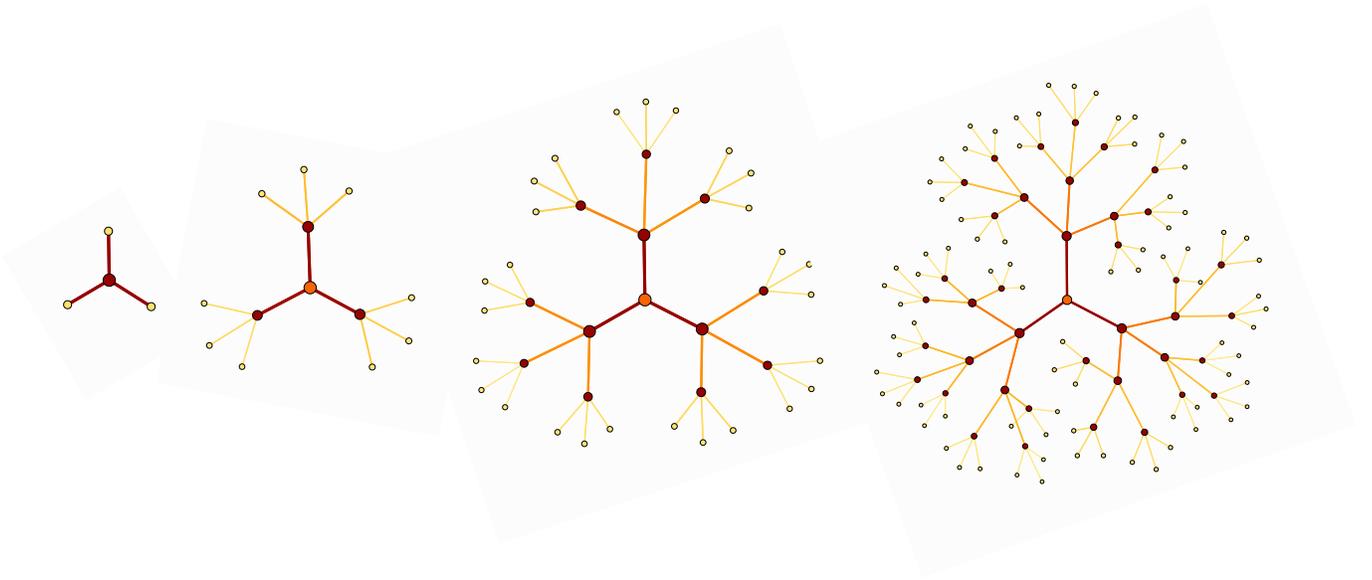}
\caption{The \lq\lq Sierpinski\rq\rq \, WFN, $s=3$, $f=1/2$ and
  $G_0$ is composed by a single node. From the left to the right $G_1$, $G_2$,
  $G_3$ and $G_4$. Gray scale (color on-line) reproduces edges weights: the
  darker the color 
  the larger the weight. The dimension of the fractal is $\log 3/\log 2\sim
  1.5850$. Visualization was done using Himmeli software~\cite{Himmeli}.} 
\label{fig:grafo1}
\end{figure*}

\begin{figure*}
\centering
\includegraphics[width=18cm]{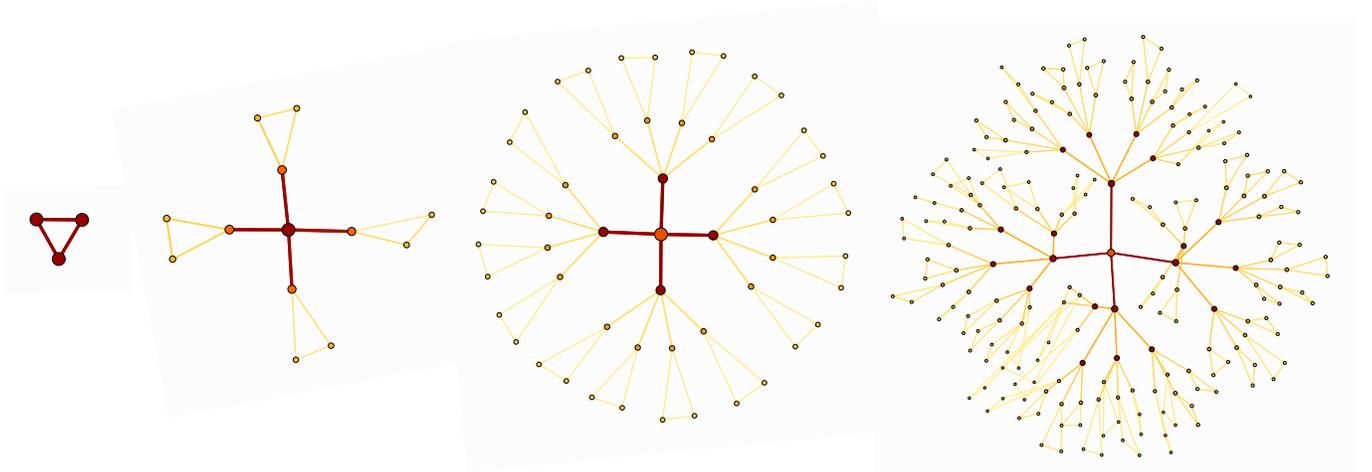}
\caption{The \lq\lq Cantor dust\rq\rq \, WFN, $s=4$, $f=1/5$ and
  $G_0$ is a triangle. From the left to the right $G_0$, $G_1$, $G_2$ 
  and $G_3$. Gray scale (color on-line) reproduces edges weights: the darker
  the color the larger the weight. The dimension of the fractal is $\log
  4/\log 5\sim 
  0.8614$. Visualization was done using Himmeli software~\cite{Himmeli}.} 
\label{fig:grafo3}
\end{figure*}

Because of its general definition, the map $\mathcal{T}_{s,f,a}$
improves the constructions recently proposed
in~\cite{Zhang2008,Zhang2008EPJB65,Guan2009}, allowing us to consider all
possible IFS fractals in a unified scheme instead of using \lq\lq ad hoc\rq\rq
constructions. For 
the sake of completeness we 
present 
numerical results for two WFN: the {\it 
  Sierpinski} one (see Fig.~\ref{fig:grafo1}) and the {\it Cantor dust} (see
Fig.~\ref{fig:grafo3}). 

Given $G_0$ and the map $\mathcal{T}_{s,f,a}$ we
are able to completely characterize the topology of each $G_k$ and also of the
limit network $G_{\infty}$, defined as the fixed point of the map:
$G_{\infty}=\mathcal{T}_{s,f,a}(G_{\infty})$. Thus the WFN undergo through a
growth 
process strictly related to the inverse of the renormalization
procedure~\cite{SongHavlinMakse2005,SongHavlinMakse2006}; at the same time
$G_{\infty}$ will be infinitely renormalizable.

\section{Results}
\label{sect:result}

 The aim of this section is to characterize the topology of the graphs $G_k$ for
 all $k\geq 1$ and $G_{\infty}$, by analytically
studying their properties such as the average degree, the node
strength distribution, the average (weighted) shortest path and the average
(weighted) clustering coefficient.  

At each iteration step the graph $G_k$ grows as the number of its nodes
increases according to 
\begin{equation}
  \label{eq:numnodes}
  N_k=s^kN_0+(s^{k}-1)/(s-1)\, ,
\end{equation}
being $N_0$ the
number of nodes in the initial graph, while the number of edges satisfies
\begin{equation}
  \label{eq:numedges}
  E_k=s^kE_0+s(s^{k}-1)/(s-1)\, ,
\end{equation}
being $E_0$ the number of edges in the graph $G_0$. Hence in the limit of
large $k$ the {\it average degree} is finite and it is asymptotically given by 
\begin{equation}
  \label{eq:asymtavdeg}
  \frac{E_k}{N_k}\underset{k\rightarrow \infty}{\longrightarrow}\frac{s+E_0(s-1)}{1+(s-1)N_0}\, .
\end{equation}

Let us denote the weighted degree of
node $i\in G_k$, also called {\it node strength}~\cite{BBPV2004}, by
$\omega^{(k)}_i=\sum_{j}w_{ij}^{(k)}$, being $w_{ij}^{(k)}$ the weight
of the edge $(ij)\in G_k$; then using the recursive construction,
we can explicitly compute the total node 
strength, $W_k=\sum_{i}\omega^{(k)}_{i}$, and, provided $sf\neq 1$, easily show
that $$W_k=2s\frac{(sf)^k-1}{sf-1}+(sf)^kW_0\, .$$ Because $f<1$, we
trivially find that the {\it average node strength} goes to zero as $k$
increases: 
${W_k}/{N_k}\underset{k\rightarrow \infty}{\longrightarrow} 0$.

\subsection{Node strength distribution.}
\label{ssec:degdist}

\begin{figure}
\centering
\includegraphics[width=9cm]{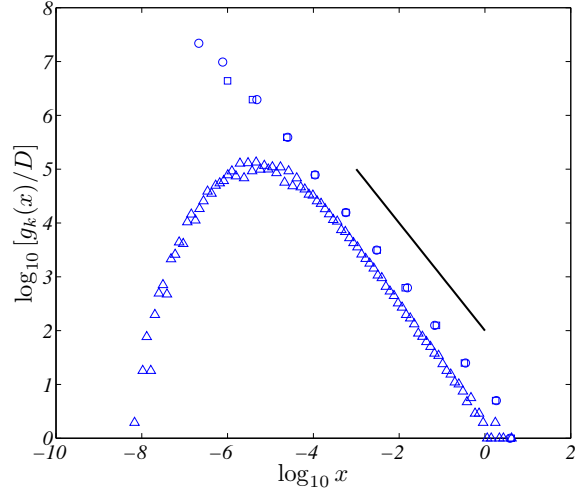}
\caption{Node Strengths Distribution. Plot of the renormalized node strengths
  distribution $\log_{10} \left[g_k(x)/D\right]$, where $D=d_{frac}$ in the
  homogeneous 
  case, while 
  $D=-\frac{s\log s}{\log (f_1\dots f_s)}$ in the non--homogeneous one.
Symbols refer to :
  $\Box$ the finite approximation $G_{14}$ with $2391484$ nodes of the
  \lq\lq Sierpinski\rq\rq \, WFN, $s=3$, $f=1/2$ and 
  $G_0$ is formed by one
  initial node; $\bigcirc$ the finite approximation $G_{11}$ composed by
  $3495253$ nodes of the \lq\lq Cantor dust\rq\rq \, WFN, $s=4$,
  $f=1/5$ and 
  $G_0$ is made by a triangle; $\bigtriangleup$ the finite approximation
  $G_{11}$ composed by $3495253$ nodes of the non--homogeneous \lq\lq
  Cantor dust\rq\rq \, WFN, $s=4$, $f_1=1/2$, $f_2=1/3$, $f_3=1/5$, $f_4=1/7$
  and $G_0$ is formed by a triangle.
The reference line has slope $-1$; linear best fits  (data not shown) provide
a slope $-0.9964\pm 0.034$ with $R^2=0.9993$ for the Sierpinski WFN, a slope
$-1.002\pm 0.064$ with $R^2=0.9996$ for the Cantor dust WFN and a slope
$-1.006\pm 0.024$ with $R^2=0.9976$ for the non--homogeneous Cantor dust WFN.}
\label{fig:construction1bis}
\end{figure}

Let $g_k(x)$ denote the number of nodes in $G_k$ that have strength
$\omega^{(k)}_i=x$ and let us assume $g_0$ to have values in some finite 
discrete subset of the positive reals, namely: 
\begin{equation*}
g_0(x)>0 \; \text{if and
  only if} \; x\in\{x_1,\dots,x_m\}\, ,
\end{equation*}
otherwise $g_0(x)=0$. Using the property of the map $\mathcal{T}_{s,f,a}$ we
straightforwardly get 
$g_k(x)=sg_{k-1}(x/f)$ provided~\footnote{  
Without loose of generality we can assume that for all integers $i,j\in\{
1,\dots, m \}$ and
$k>0$ we have $f^kx_j\neq x_i$ and $f^k(fx_j+1)\neq x_i$.}
$x\neq s$ and $x\neq fs+1$, from which we can conclude that for all $k$:
\begin{equation}
  \label{eq:degdist}
  g_k(x)=s^kg_{0}(x/f^k)\, , \quad g_k(fs+1)=s\quad  \text{and}\quad g_k(s)=1\, .
\end{equation}
This implies than the node strengths are distributed according to a power
law with exponent $d_{fract}=-\log s/\log f$, that equals the fractal
dimension of the 
fractal obtained as fixed point of the IFS with the same parameters $s$ and
$f$. In 
fact defining 
$x_{ik}=f^k x_i$ we get:
\begin{eqnarray*}
  \log g_k(x_{ik})&=&k\log s +\log g_0(x_i)\\ &=&\frac{\log s}{\log f}\log
  x_{ik}+\log g_0(x_i)-\frac{\log s}{\log f}\log x_{i}\, ,
\end{eqnarray*} 
namely (see Fig.~\ref{fig:construction1bis})
\begin{equation}
  \label{eq:nodestrengths}
  g_k(x)\sim C/x^{d_{frac}}\, .
\end{equation}

\begin{figure}
\centering
\includegraphics[width=9cm]{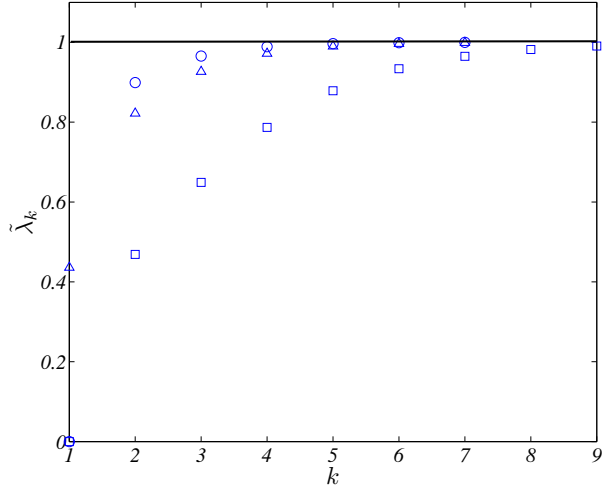}
\caption{The average weighted shortest path. Plot of the renormalized average
  weighted shortest 
  path $\tilde{\lambda}_k$ versus the iteration
  number $k$, where
  $\tilde{\lambda}_k=\lambda_k\frac{(s-F)(s^2-F)}{2s^2(s-1)}$ and
  $F=f_1+\dots+f_s$ for the non--homogeneous case, while $F=sf$ for the
  homogeneous one. Symbols refer to :
  $\Box$ the \lq\lq Sierpinski\rq\rq \, WFN, $s=3$, $f=1/2$ and
  $G_0$ is formed by one
  initial node; $\bigcirc$ the \lq\lq Cantor dust\rq\rq \, WFN, $s=4$,
  $f=1/5$ and 
  $G_0$ is made by a triangle; $\bigtriangleup$ the non--homogeneous \lq\lq
  Cantor dust\rq\rq \, WFN, $s=4$, $f_1=1/2$, $f_2=1/3$, $f_3=1/5$, $f_4=1/7$
  and $G_0$ is formed by a triangle.}
\label{fig:construction1ter}
\end{figure}

\subsection{Average weighted shortest path.}
\label{ssec:meanpath}

By definition the average {\it weighted shortest path}~\cite{BLMCH2006} of the
graph $G_k$ is given by
\begin{equation}
\label{eq:wmpath}
 \lambda_k=\frac{\Lambda_k}{N_k(N_k-1)}\, ,
\end{equation}
 where
\begin{equation}
\label{eq:totwmpath}
\Lambda_k=\sum_{ij\in G_k} p_{ij}^{(k)}\, ,
\end{equation}
being $p_{ij}^{(k)}$ the weighted shortest path linking nodes $i$ and $j$ in
$G_k$.  

To simplify the remaining part of the proof it is useful to introduce
$\Lambda_k^{(a_k)}=\sum_{i\in G_k}p_{ia_k}^{(k)}$, i.e. the sum of all weighted shortest
paths ending at the attaching node, $a_k\in G_k$. One can prove (see
Appendix~\ref{ssec:Lkc}) that for large $k$ the asymptotic behavior of 
$\Lambda_k^{(a_k)}$ is given by
\begin{equation}
  \label{eq:Lkcasymt}
  \Lambda_k^{(a_k)}\underset{k\rightarrow \infty}{\sim} \frac{N_0(s-1)+1}{(1-f)(s-1)}s^{k-1}\, .
\end{equation}

Using the construction algorithm and its symmetry one can prove (see
again the Appendix~\ref{ssec:Lk}) that $\Lambda_k$ satisfies the recursive
relation 
\begin{equation}
  \label{eq:Lkrec}
  \Lambda_k=sf\Lambda_{k-1}+2s[(s-1)N_{k-1}+1][N_{k-1}+f\Lambda_{k-1}^{(a_{k-1})}]\, ,
\end{equation}
that provides the following asymptotic behavior in the limit of large $k$
(see Fig.~\ref{fig:construction1ter}) 
\begin{equation}
  \label{eq:ellkasym}
  \lambda_k=\frac{\Lambda_k}{N_k(N_k-1)}\underset{k\rightarrow \infty}{\longrightarrow }
  \frac{2(s-1)}{(1-f)(s-f)}\, . 
\end{equation}

\begin{figure}
\centering
\includegraphics[width=9cm]{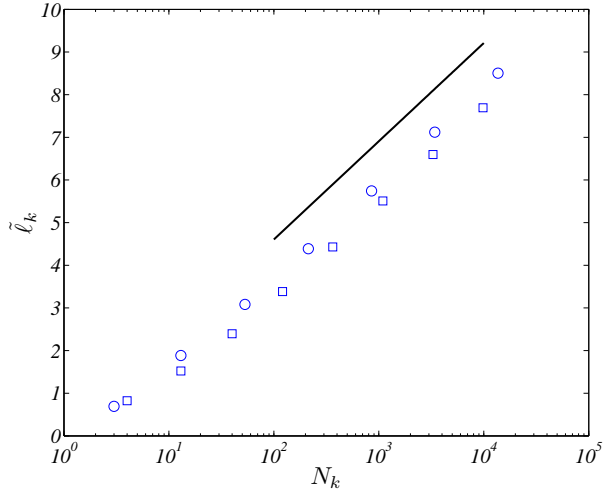}
\caption{The average shortest path $\ell_k$ as a function of the network size
  (semilog graph). Plot of the renormalized average shortest
  path $\tilde{\ell}_k$ versus the network size $N_k$, where
  $\tilde{\ell}_k=\ell_k \frac{\log s}{2}$. Symbols refer to :
  $\Box$ the \lq\lq Sierpinski\rq\rq \, WFN, $s=3$, $f=1/2$ and
  $G_0$ is formed by one
  initial node; $\bigcirc$ the \lq\lq Cantor dust\rq\rq \, WFN, $s=4$, $f=1/5$
  and 
  $G_0$ is made by a triangle. The reference line has slope $1$. Linear best
  fits (data not shown) provides a slope $0.9942\pm 0.019$ and
  $R^2=1$ for the Sierpinski WFN and a slope $0.9952\pm 0.019$ and
  $R^2=1$ for the Cantor dust WFN.}  
\label{fig:ellek}
\end{figure}

We can also compute the {\it average shortest path}, $\ell_k$, formally
obtained by setting $f=1$ in the previous formulas~\eqref{eq:wmpath}
and~\eqref{eq:totwmpath}. Hence slightly 
modifying the results previously presented we can prove that asymptotically
we have 
\begin{equation}
  \label{eq:avmeanpath}
  \ell_k \underset{k\rightarrow \infty}{\sim}2\left(
    k-\frac{s}{s-1}\right)\underset{k\rightarrow \infty}{\sim} \frac{2}{\log
    s}\log N_k\, ,
\end{equation}
where the last relation has been obtained using the growth law of $N_k$ given
by equation~\eqref{eq:numnodes} (see Fig.~\ref{fig:ellek}). Let us remark that
the average shortest path is a topological quantity and thus it doesn't depend
on the scaling factor, that's why we don't report in Fig.~\ref{fig:ellek} the
case of the non-homogeneous WFN.

Thus, as previously stated, the
network grows unbounded but with 
the logarithm of the network size, while the weighted shortest distances 
stay bounded.

\subsection{Average clustering coefficient.}
\label{ssec:cluscoeff}

The average clustering coefficient~\cite{WattsStrogatz1998,BLMCH2006} of the
graph 
$G_k$ is defined as the average over the whole set of
nodes of the local clustering coefficient $c_i^{(k)}$, namely $<c_k>=C_k/N_k$,
where $C_k=\sum_{i\in G_k}c^{(k)}_i$. Because of the 
construction algorithm the number of possible triangles, hence the local
clustering 
coefficient, at each 
step increases by a factor $s$; thus after $k$ iterations we will have
$C_k=s^{k}C_0$, being $C_0$ the sum of
local clustering coefficients in the initial graph. We can thus conclude that
the clustering 
coefficient of the graph is asymptotically 
given by: 
\begin{equation}
  \label{eq:asymptclustcoeff}
  <c_k>\underset{k\rightarrow \infty}{\longrightarrow }
  \frac{s-1}{s}\frac{<c_0>N_0}{(s-1)N_0+1}\, .  
\end{equation}

On the other hand, one can use edges' values to weigh the clustering
coefficient~\cite{SKOKK2007}; hence generalizing the previous relation, we can
easily prove that the average {\it weighted clustering coefficient} of the
graph is 
asymptotically given by: 
\begin{equation}
  \label{eq:asymptclustcoeffweig}
  <\gamma_k>\underset{k\rightarrow \infty}{\sim}
  \frac{s-1}{fs}\frac{<\gamma_0>N_0}{(s-1)N_0+1}f^k \underset{k\rightarrow
    \infty}{\sim} 
  \frac{1}{N_k^{1/d_{fract}}}\, ,
\end{equation}
where once again, the fractal dimension $d_{fract}$ of the IFS fractal play a
relevant role.

\section{Non--homogeneous Weighted Fractal Networks}
\label{sec:nhwfn}

The aim of this section is to slightly generalize the previous construction to
the case of {\it non--homogeneous} scaling factors for each subnetwork
$G^{(i)}$. So given an integer $s>1$ and $s$ real numbers $f_1,\dots,f_s\in
(0,1)$, we modify the map $\mathcal{T}_{s,f,a}$ by allowing a different
scaling for each edge weight according to which subgraph it belongs to: if
the edge $w^{(j)}$, image of $w\in G$, belongs to $G^{(j)}$, then
$w^{(j)}=f_jw$. 

\begin{figure}
\centering
\includegraphics[width=18cm]{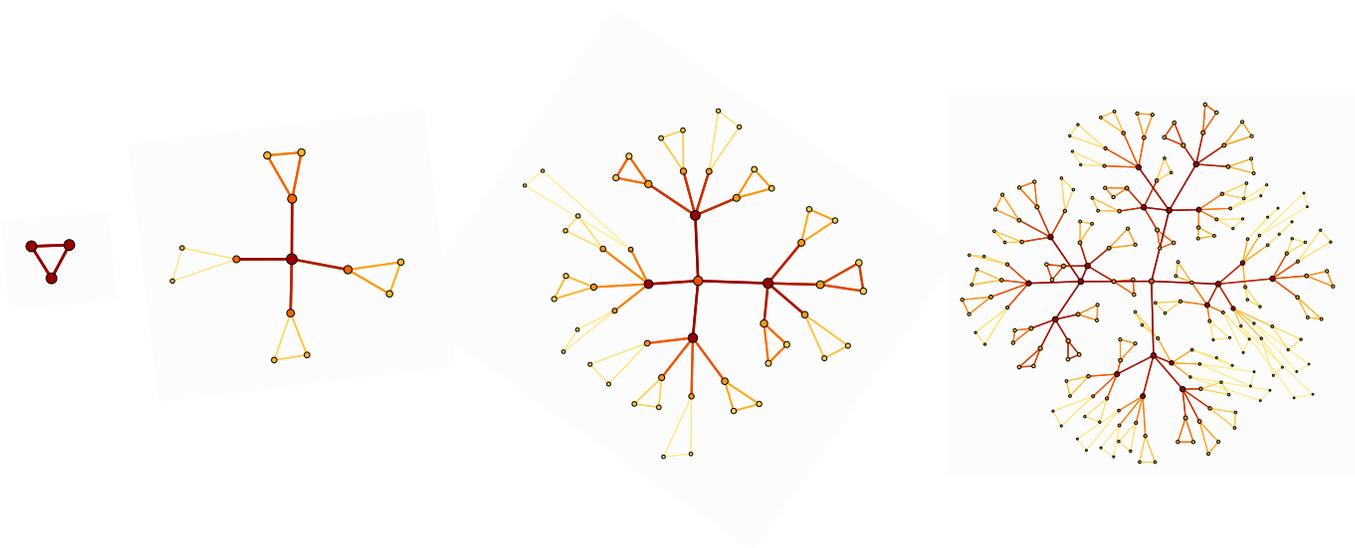}
\caption{The non--homogeneous \lq\lq Cantor dust\rq\rq \, WFN, $s=4$,
  $f_1=1/2$, $f_2=1/3$, $f_3=1/5$, $f_4=1/7$ and $G_0$ is formed by a
  triangle. From the left to the right $G_0$, $G_1$, $G_2$  
  and $G_3$. Gray scale (color on-line) reproduces edges weights: the darker
  the color the larger the weight. Visualization was done using Himmeli
  software~\cite{Himmeli}.}  
\label{fig:nhcantordust}
\end{figure}

Let us remark that the construction presented in the former Section~\ref{sect:model} is a 
particular case of the latter 
once we take $f_1=\dots=f_s=f$; we nevertheless decided for a sake of
clarity, to present it before, because the computations involved in this
latter general construction could 
have hidden the simplicity of the underlying idea. We hereby present some
results for the non--homogeneous \lq\lq Cantor dust\rq\rq WFN (see
Fig.~\ref{fig:nhcantordust}).

Using the recursiveness of the algorithm we can, once again, completely
characterize the topology of the non-homogeneous WFN, moreover only the
weighted quantities will vary with respect to the homogeneous case. For
instance, a straightforward, but cumbersome, generalization of 
the computations 
presented in the previous Sections allows us to prove that the average
weighted shortest path 
exhibits the following asymptotic behavior (see
Fig.~\ref{fig:construction1ter})  
\begin{equation}
  \label{eq:ellkasymnh}
  \lambda_k\underset{k\rightarrow \infty}{\longrightarrow }
  \frac{2s^2(s-1)}{(s-F)(s^2-F)}\, ,
\end{equation}
where $F=f_1+\dots+f_s$. Let us observe that Eq.~\eqref{eq:ellkasymnh} reduces
to Eq.~\eqref{eq:ellkasym} once we set $f_1=\dots=f_s=f$ and thus $F=sf$.

Let $g_0(x)$ denote the number of nodes with node strength equal to $x$ in the
initial network $G_0$; then after $k$ steps of the algorithm, all nodes
strengths will be rescaled by a factor $f_1^{k_1}\dots f_s^{k_s}$, where the
non-negative integers $k_i$ do satisfy $k_1+\dots +k_s=k$. Because this can be
done in $k!/(k_1!\dots k_s!)$ possible different ways, we get the following
relation for the node strength distribution for the network $G_k$:
\begin{equation}
  \label{eq:nhnodestrength}
  g_k(f_1^{k_1}\dots f_s^{k_s}x)=\frac{k!}{k_1!\dots k_s!}g_0(x)\quad
  \text{with $k_1+\dots+k_s=k$}\, .
\end{equation}
After sufficiently many steps and assuming that the main contribution arises
from the choice $k_1\sim \dots \sim k_s \sim k/s$, we can use Stirling formula
to get the approximate distribution (see Fig.~\ref{fig:construction1bis})
\begin{equation}
  \label{eq:nhnodestrapprx}
  \log g_k(x) \sim \frac{s\log s}{\log (f_1\dots f_s)} \log x\, ,
\end{equation}
so once again the nodes strength distribution follows a power law.

\section{Conclusions}
\label{sect:conclusion}

In this paper we introduced a unifying framework for complex networks sharing
several properties with fractal sets, hereby named {\it Weighted Fractal
Networks}. This theory, that generalizes to graphs the construction of
IFS fractals, allows us to build complex networks
with a prescribed topology, whose main quantities can be analytically
predicted and have been shown to depend on the fractal dimension of the IFS
fractal; for instance the networks are scale--free with exponent the
fractal dimension. Moreover the weighted fractal networks share with IFS fractals, the
self-similarity structure, and are explicitely computable examples of
renormalizable complex networks.

These networks exhibit the {\it small--world}
property. In 
fact the average shortest path increases logarithmically with the system
size~\eqref{eq:avmeanpath}, hence it is small as the average shortest path of a
random network with the same number of nodes and same average degree. On the
other hand the clustering coefficient is asymptotically
constant~\eqref{eq:asymptclustcoeff}, thus larger than the clustering
coefficient of a random network that shrinks to zero as the system size
increases.

The self-similarity property of the weighted fractal networks makes them
suitable to model real problems involving generic diffusion over the
network coupled with local looses of flow, here modeled via the parameter
$f<1$. For instance one can think of electrical grids or mammalian 
lungs, where current or air, flows through power
lines or bronchi--bronchioles, submitted to looses of
power, or air vessels' section reduction. In all these cases the
induced topology, namely a good choice of $f$ and $s$, allows any two random
nodes, final power users or alveoli, to be always at finite weighted distance,
whatever their physical 
distance is, and thus to be able to transport current or oxygen in finite time.


\appendix
\section{Complementary material}
\label{sect:app}

\subsection{Computation of $\Lambda_k^{(a_k)}$}
\label{ssec:Lkc}

Let $a_k$ be the attaching node of the graph $G_k$. Let us define
$\Lambda_k^{(a_k)}=\sum_{i\in G_k}p_{ia_k}^{(k)}$, i.e. the sum of all weighted
shortest paths to $a_k$. Then using the
recursive property and the symmetry of the map $\mathcal{T}_{s,f,a_k}$  we
can easily obtain a recursive relation for
$\Lambda_k^{(a_k)}$: $$\Lambda_k^{(a_k)}=sf\Lambda_{k-1}^{(a_{k-1})}+sN_{k-1}\,
,$$ where $N_{k-1}$ is the number of nodes in $G_{k-1}$. This recursion can be
easily solved to get for all $k\geq 1$
\begin{eqnarray}
  \label{eq:Lkc}
  \Lambda_k^{(a_k)}=(sf)^{k-1}\Lambda_1^{(a_1)}+\frac{1-f^{k-1}}{1-f}\frac{(s-1)N_0+1}{s-1}
  s^{k-1}-\frac{s}{s-1}\frac{(sf)^{k-1}-1}{sf-1}\, ,  
\end{eqnarray}
from which we can conclude, because $f<1$, that $\Lambda_k^{(a_k)}$ exhibits
the asymptotic behavior given by equation~\eqref{eq:Lkcasymt}.

\subsection{Computation of $\Lambda_k$}
\label{ssec:Lk}

Starting from the definition of the sum of all weighted shortest
paths~\eqref{eq:totwmpath}, the recursive construction and its symmetry we
can decompose the sum $\Lambda_k$ into three terms: 
\begin{equation}
  \label{eq:Lkinter}
\Lambda_k=s\sum_{ij\in G_k^{(1)}}p_{ij}^{(k)}+s(s-1)\sum_{i\in G_k^{(1)},j\in
      G_k^{(2)}}p_{ij}^{(k)}+2s\sum_{i\in G_k^{(1)}}p_{ia_k}^{(k)}
\end{equation}
where the first contribution takes into account all paths starting from and
arriving to nodes belonging to the same subgraph, that using the symmetry can be
chosen to be $G_k^{(1)}$. The second term takes 
into account all the possible paths where the initial point and the final one
belong to two different subgraphs, and still using the symmetry we can set
them to $G_k^{(1)}$ and $G_k^{(2)}$ and multiply the contribution by a
combinatorial factor
$s(s-1)$. Finally the last term is the sum of 
all paths arriving to the attaching node $a_k$; once again the symmetry allows
us to reduce the sum to only one subgraph, say $G_k^{(1)}$, and multiply the
contribution by $2s$.

Using the scaling mechanism for the edges, the first term in the right hand
side of 
equation~\eqref{eq:Lkinter} can be 
easily 
identified with 
\begin{equation*}
\sum_{ij\in G_k^{(1)}}p_{ij}^{(k)}=f\Lambda_{k-1}\, .
\end{equation*}

By construction, each shortest path connecting two nodes belonging to two
different subgraphs, must pass through the attaching node, hence using
$p_{ij}^{(k)}=p_{ia_k}^{(k)}+p_{a_kj}^{(k)}$ the second term
of equation~\eqref{eq:Lkinter} can be split into two parts:
\begin{equation*}
\sum_{i\in G_k^{(1)},j\in G_k^{(2)}}p_{ij}^{(k)}=\sum_{i\in
  G_k^{(1)}}p_{ia_k}^{(k)}N_k^{(2)}+\sum_{j\in
  G_k^{(2)}}p_{a_kj}^{(k)}N_k^{(1)}\, ,
\end{equation*}
where $N_k^{(i)}$ denotes the number of nodes in the subgraph $G^{(i)}_k$. Using
the symmetry of the construction, the previous relation can be rewritten as
\begin{equation*}
\sum_{i\in G_k^{(1)},j\in G_k^{(2)}}p_{ij}^{(k)}=2N_k^{(1)}\sum_{i\in
  G_k^{(1)}}p_{ia_k}^{(k)}\, .
\end{equation*}

The last term of equation~\eqref{eq:Lkinter} can be related to
$\Lambda_{k-1}^{(a_{k-1})}$ 
by observing that each path arriving at $a_k$ must pass through
$a_k^{(i)}$ for some $i\in \{1,\dots,s\}$, thus
\begin{eqnarray}
\sum_{i\in  G_k^{(1)}}p_{ia_k}^{(k)}&=&\sum_{i\in
  G_k^{(1)}}(p_{ia^{(1)}_k}^{(k)}+p_{a^{(1)}_ka_k}^{(k)})=N_k^{(1)}+\sum_{i\in
  G_k^{(1)}} p_{ia^{(1)}_k}^{(k)}\notag \\
&=&N_k^{(1)}+f\Lambda_{k-1}^{(a_{k-1})}\, , 
\end{eqnarray}
Observing that $G_k^{(1)}$ has as many nodes as $G_{k-1}$ we can
conclude that $N_k^{(1)}=N_{k-1}$ and finally to rewrite
equation~\eqref{eq:Lkinter} as: 
\begin{equation*}
\Lambda_k=sf\Lambda_{k-1}+2s[(s-1)N_{k-1}+1][N_{k-1}+f\Lambda_{k-1}^{(a_{k-1})}]\, .
\end{equation*}

\end{document}